\documentclass[12pt]{article}
\usepackage{amsmath,amssymb,epsfig}
\begin{document}
\parindent=0pt
\parskip=6pt
\rm

\vspace{0.5cm}

\small

\begin{center}
{\bf \large New fluctuation-driven phase transitions and critical
phenomena in unconventional superconductors}

\vspace{0.5cm}

 { \large Dimo I. Uzunov}

 CP Laboratory,
 Institute of Solid State Physics, Bulgarian Academy of Sciences,\\ BG--1784, Sofia, Bulgaria.

\end{center}

\begin{abstract} Using the renormalization group method, new type of fluctuation-driven
first order phase transitions and critical phenomena are predicted
for certain classes of ferromagnetic superconductors and
superfluids with unconventional (spin-triplet) Cooper pairing. The
problem for the quantum phase transitions at extremely low and
zero temperatures is also discussed. The results can be applied to
a wide class of ferromagnetic superconductive and superfluid
systems, in particular, to itinerant ferromagnets as UGe$_2$ and
URhGe.
\end{abstract}

{\bf Pacs:} 05.70.Jk,74.20.De, 75.40.Cx

\vspace{0.5cm}

{\bf 1. Introduction}

 In this paper an entirely new critical
behavior in unconventional ferromagnetic superconductors and
superfluids is established and described. This phenomenon
corresponds to an isotropic ferromagnetic order in real systems
but does not belong to any known universality
class~\cite{Uzunov:1993} and, hence, it might be of considerable
experimental and theoretical interest. Due to crystal and magnetic
anisotropy a new type of fluctuation-driven first order phase
transitions occur, as shown in the present investigation. These
novel fluctuation effects can be observed near finite and zero
temperature (``quantum'') phase transitions~\cite{Uzunov:1993,
ShopovaPR:2003} in a wide class of ferromagnetic systems with
unconventional (spin-triplet) superconductivity or superfluidity.

The present investigation has been performed on the concrete
example of intermetallic compounds UGe$_2$ and URhGe, where the
remarkable phenomenon of coexistence of itinerant ferromagnetism
and unconventional spin-triplet
superconductivity~\cite{Sigrist:1991} has been
observed~\cite{Saxena:2000}. For example, in UGe$_2$, the
coexistence phase occurs~\cite{Saxena:2000} at temperatures $0\leq
T <1$~K and pressures $1<P < P_0 \sim 1.7$~GPa. A fragment of
($P,T$) phase diagrams of itinerant ferromagnetic
compounds~\cite{Saxena:2000} is sketched in Fig.~1, where the
lines $T_F(P)$ and $T_c(P)$ of the paramagnetic(P)
-to-ferromagnetic(F) and ferromagnetic-to-coexistence phase(C)
transitions are very close to each other and intersect at very low
temperature or terminate at the absolute zero ($P_0,0$). At low
temperature, where the phase transition lines are close enough to
each other, the interaction between the real magnetization vector
$\mbox{\boldmath$M$}(\mbox{\boldmath$r$)} =
\{M_j(\mbox{\boldmath$r$}); j=1,...,m \}$ and the complex order
parameter vector of the spin-triplet Cooper
pairing~\cite{Sigrist:1991}, $\psi(\mbox{\boldmath$r$}) =
\{\psi_{\alpha}(\mbox{\boldmath$r$}) = (\psi^{\prime}_{\alpha} +
i\psi_{\alpha}^{\prime \prime}); \alpha =1,.... n/2\}$ ($n=6$)
cannot be neglected~\cite{Uzunov:1993} and, as shown here, this
interaction produces new fluctuation phenomena.
\begin{figure}
\begin{center}
\epsfig{file=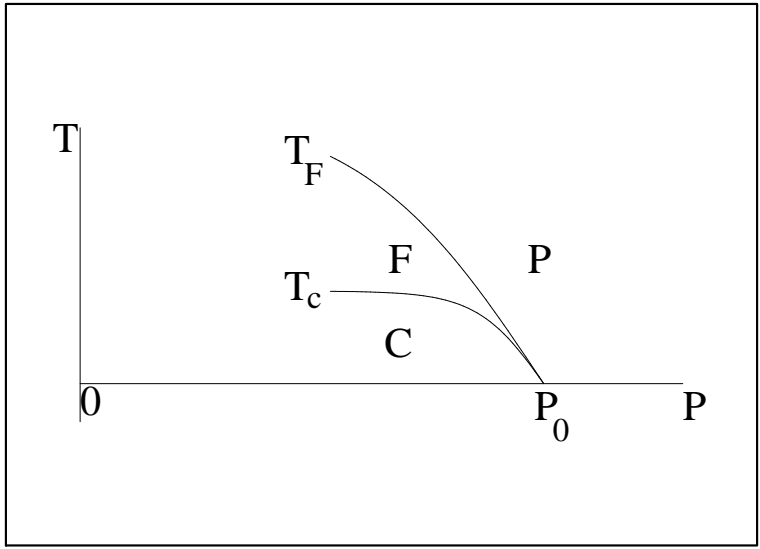,angle=-90, width=10cm}\\
\end{center}
\caption{\footnotesize ($P,T$) diagram with a zero-temperature
multicritical point $(P_0,0)$. Para- (P), ferromagnetic (F), and
coexistence (C) phases, separated by the lines $T_f(P)$ and
$T_c(P)$ of P-F and F-C phase transitions, respectively.}
\label{NEW1.fig}
\end{figure}

Both thermal fluctuations at finite temperatures ($T>0$) and
quantum fluctuations (correlations) near the $P$--driven quantum
phase transition at $T=0$ should be considered but at a first
stage the quantum effects~\cite{ShopovaPR:2003} can be neglected
as irrelevant to finite temperature phase transitions ($T_F \sim
T_c
>0$). The present treatment of a recently derived free energy
functional~\cite{Machida:2001} by the standard Wilson-Fisher
renormalization group (RG)~\cite{Uzunov:1993} shows that
unconventional ferromagnetic superconductors with an isotropic
magnetic order ($m=3$) exhibit a quite particular multi-critical
behavior for any $T> 0$, whereas the magnetic anisotropy ($m
=1,2$) generates fluctuation-driven first order
transitions~\cite{Uzunov:1993}. Thus the phase transition
properties of spin-triplet ferromagnetic superconductors are
completely different from those predicted by mean field
theories~\cite{Machida:2001, Shopova:2003}. The results can be
used in the interpretation of experimental data for phase
transitions in itinerant ferromagnetic
compounds~\cite{Pfleiderer:2002}.

The study presents for the first time an example of complex
quantum criticality characterized by a double-rate quantum
critical dynamics. In the quantum limit ($T\rightarrow 0$) the
fields $\mbox{\boldmath$M$}$ and $\psi$ have different dynamical
exponents, $z_M$ and $z_{\psi}$, and this leads to two different
upper critical dimensions: $d_U^M = 6-z_M$ and $d_{\psi}^U =
6-z_{\psi}$. The complete consideration of the quantum
fluctuations of both fields $\mbox{\boldmath$M$}$ and $\psi$
requires a new RG approach in which one should either consider the
difference $(z_M-z_{\psi})$ as an auxiliary small parameter or
create a completely new theoretical paradigm of description. The
considered problem is quite general and presents a challenge to
the theory of quantum phase transitions~\cite{ShopovaPR:2003}. The
results can be applied to any natural system within the same class
of symmetry although this report is based on the example of
itinerant ferromagnetic compounds.

{\bf 2. Renormalization-group investigation}

The relevant part of the fluctuation Hamiltonian of unconventional
ferromagnetic superconductors~\cite{Machida:2001,Shopova:2003} can
be written in the form
\begin{equation}
\label{eq1} {\cal{H}}= \sum_{\mbox{\boldmath$k$}}\left[ \left(r +
k^2 \right)|\psi(\mbox{\boldmath$k$})|^2 + \frac{1}{2}\left(t +
k^2\right)|\mbox{\boldmath$M$}(\mbox{\boldmath$k$})|^2   \right]
+\frac{ig}{\sqrt{V}}\sum_{\mbox{\boldmath$k$}_1,{\mbox{\boldmath$k$}_2}}{\mbox{\boldmath$M$}}
\left(
{\mbox{\boldmath$k$}}_1\right).\left[\psi\left({\mbox{\boldmath$k$}}_2
\right)\times \psi^{\ast}\left({\mbox{\boldmath$k$}}_1 +
{\mbox{\boldmath$k$}}_2 \right) \right]
\end{equation}
where $V \sim L^d$ is the volume of the $d-$dimensional system,
the length unit is chosen so that the wave vector
${\mbox{\boldmath$k$}}$ is confined below unity ($ 0 \leq k =
|{\mbox{\boldmath$k$}}| \leq 1)$, $g \geq 0$ is a coupling
constant, describing the effect the scalar product of
${\mbox{\boldmath$M$}}$ and the vector product
$(\psi\times\psi^{\ast})$ for symmetry indices $m = (n/2)=3$, and
the parameters $t \sim (T-T_f)$ and $r \sim (T-T_s)$ are expressed
by the critical temperatures of the generic ($g\equiv 0$)
ferromagnetic and superconducting transitions. As mean field
studies indicate~\cite{Machida:2001,Shopova:2003}, $T_s(P)$ is
much lower than $T_c(T)$ and $T_F(P) \neq T_f(P)$.

The fourth order terms ($M^4, |\psi|^4, M^2|\psi|^2$) in the total
free energy ({\em effective} Hamiltonian)~\cite{Machida:2001,
Shopova:2003} have not been included in Eq.~(1) as they are
irrelevant to the present investigation. The simple dimensional
analysis shows that the $g-$term in Eq.~(1) corresponds to a
scaling factor $b^{3-d/2}$ and, hence, becomes relevant below the
upper borderline dimension $d_U=6$, while fourth order terms are
scaled by a factor $b^{4-d}$ as in the usual $\phi^4-$theory and
are relevant below $d<4$ ($b > 1$ is a scaling
number)~\cite{Uzunov:1993}. Therefore we should perform the RG
investigation in spatial dimensions $d = 6-\epsilon$ where the
$g$--term in Eq.~(1) describes the only relevant fluctuation
interaction. Moreover, the total fluctuation
Hamiltonian~\cite{Machida:2001, Shopova:2003} contains
off-diagonal terms of the form
$k_ik_j\psi_{\alpha}\psi^{\ast}_{\beta}$; $i\neq j$ and/or $\alpha
\neq \beta$. Using a convenient loop expansion these terms can be
completely integrated out from the partition function to show that
they modify the parameters ($r,t,g$) of the theory but they do not
affect the structure of the model (1). So, such terms change
auxiliary quantities, for example, the coordinates of the RG fixed
points (FPs) but they do not affect the main RG results for the
stability of the FPs and the values of the critical exponents.
Here we ignore these off-diagonal terms.

One may consider several cases: (i) uniaxial magnetic symmetry,
${\mbox{\boldmath$M$}} = (0,0,M_3)$, (ii) tetragonal crystal
symmetry when $\psi = (\psi_1,\psi_2,0)$, (iii) {\em XY} magnetic
order $(M_1,M_2,0)$, and (iv) the general case of cubic crystal
symmetry and isotropic magnetic order ($m=3$) when all components
of the three dimensional vectors ${\mbox{\boldmath$M$}}$ and
$\psi$ may have nonzero equilibrium and fluctuation components.
 The latter case is of major interest to real systems where
fluctuations of all components of the fields are possible despite
the presence of spatial crystal and magnetic anisotropy that
nullifies some of the equilibrium field components. In one-loop
approximation, the RG analysis reveals different pictures for
anisotropic (i)-(iii) and isotropic (iv) systems. As usual, a
Gaussian (``trivial'') FP ($g^{\ast} =0$) exists for all $d>0$
and, as usual~\cite{Uzunov:1993} this FP is stable for $d>6$ where
the fluctuations are irrelevant. In the reminder of this paper the
attention will be focussed on spatial dimensions $d < 6$, where
the critical behavior is usually governed by nontrivial FPs
($g^{\ast} \neq 0$). In the cases (i)-(iii) only negative
(``unphysical''~\cite{Lawrie:1987}) FP values of $g^2$ have been
obtained for $d<6$. For example, in the case (i) the RG relation
for $g$ takes the form
\begin{equation}
\label{eq2} g^{\prime}= b^{3-d/2-\eta}g\left(1 +
g^2K_d{\mbox{ln}}b\right),
\end{equation}
where $g^{\prime}$ is the renormalized value of $g$, $\eta =
(K_{d-1}/8)g^2$ is the anomalous dimension (Fisher's
exponent)~\cite{Uzunov:1993} of the field $M_3$; $K_d =
2^{1-d}\pi^{-d/2}/\Gamma(d/2)$. Using Eq.~(2) one obtains the FP
coordinate $(g^2)^{\ast} = -96\pi^3\epsilon$. For $d < 6$ this FP
is unphysical and does not describe any critical behavior. For $d
> 6$ the same FP is physical but unstable towards the parameter
$g$ as one may see from the positive value $y_g = -11\epsilon /2
>0$ of the respective stability exponent
$y_g$ defined by $\delta g^{\prime} = b^{y_g}\delta g$. Therefore,
a change of the order of the phase transition from second order in
mean-field (``fluctuation free'') approximation to a
fluctuation-driven first order transition when the fluctuation
$g$--interaction is taken into account takes place. This
conclusion is supported by general concepts of RG
theory~\cite{Uzunov:1993} and by the particular property of these
systems to exhibit first order phase
transitions~\cite{Shopova:2003} in mean field approximation for
broad variations of $T$ and $P$.

In the case (iv) of isotropic systems the RG equation for $g$ is
degenerate and the $\epsilon$-expansion breaks down. A similar
situation is known from the theory of disordered
systems~\cite{Lawrie:1987} but here the physical mechanism and
details of description are different. Namely for this degeneration
one should consider the RG equations up to the two-loop order. The
derivation of the two-loop terms in the RG equations is quite
nontrivial because of the special symmetry properties of the
interaction $g$-term in Eq.~(1). For example, some diagrams with
opposite arrows of internal lines, as the couple shown in
Fig.~(2), have opposite signs and compensate each other. The terms
bringing contributions to the $g$--vertex are shown
diagrammatically in Fig.~3. The RG analysis is carried out by a
completely new $\epsilon^{1/4}$-expansion for the FP values and
$\epsilon^{1/2}$-expansion for the critical exponents; again
$\epsilon = (6-d)$. The RG equations are quite lengthy and here
only the equation for $g$ is discussed. It has the form
\begin{equation}
\label{eq3} g^{\prime}=
b^{(\epsilon-2\eta_{\psi}-\eta_M)/2}g\left[1 + Ag^2 +
3(2B+C)g^4\right],
\end{equation}
where
\begin{equation}
\label{eq4} A=\frac{K_d}{2}\left[2{\mbox{ln}}b + \epsilon
({\mbox{ln}}b)^2 + (1-b^2)(2r+t)\right],
\end{equation}
\begin{equation}
\label{eq5} B= \frac{K_{d-1}K_d}{192}\left[9(b^2-1) -
11{\mbox{ln}}b - 6\left({\mbox{ln}}b\right)^2\right],
\end{equation}
\begin{equation}
\label{eq6} C= \frac{3K_{d-1}K_d}{64}\left[{\mbox{ln}}b +
2\left({\mbox{ln}}b\right)^2\right],
\end{equation}
$\eta_M$ and $\eta_{\psi}$ are the anomalous dimensions of the
fields ${\mbox{\boldmath$M$}}$ and $\psi$, respectively. The
one-loop approximation gives correct results to order
$\epsilon^{1/2}$ and the two-loop approximation brings such
results up to order $\epsilon$. In Eq.~(4), $r$ and $t$ are small
expansion quantities with equal FP values $t^{\ast} = r^{\ast} =
K_d g^2$. Using the condition for invariance of the two
$k^2$-terms in Eq.~(1) one obtains $\eta_M =\eta_{\psi} \equiv
\eta$, where
\begin{equation}
\label{eq7} \eta =\frac{K_{d-1}}{8}g^2\left(1-
\frac{13}{96}K_{d-1}g^2\right).
\end{equation}
Eq.~(3) yields a new FP
\begin{equation} \label{eq8}
g^{\ast}=8\left(3\pi^3\right)^{1/2}\left(2\epsilon/13\right)^{1/4},
\end{equation}
which corresponds to the critical exponent $\eta =
2(2\epsilon/13)^{1/2} - 2\epsilon/3$ (for $d=3$, $\eta \approx
-0.64 $).
\begin{figure}
\begin{center}
\epsfig{file=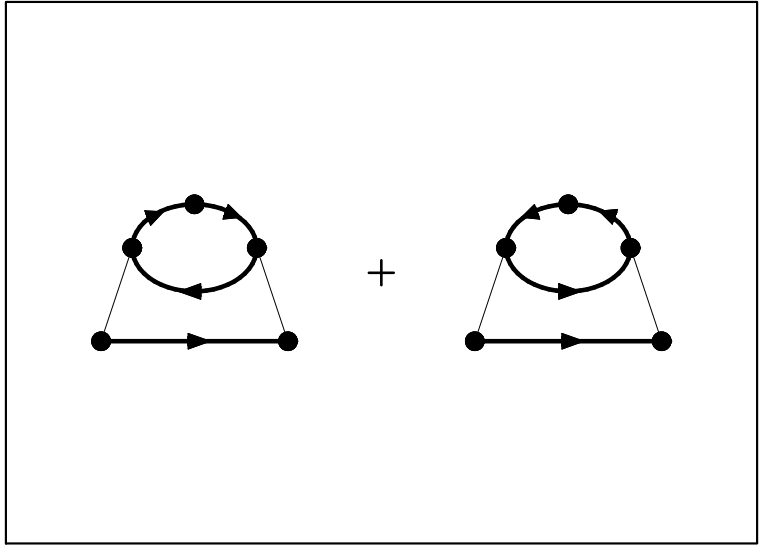,angle=-90, width=10cm}\\
\end{center}
\caption{\footnotesize A sum of $g^5$--diagrams equal to zero. The
thick and thin lines correspond to correlation functions $\langle
|\psi_{\alpha}|^2\rangle$ and $\langle |M_j|^2\rangle$,
respectively; vertices ($\bullet$) represent $g$--term in Eq.~(1).
} \label{NEW1.fig}
\end{figure}

The eigenvalue problem for the RG stability matrix
$\hat{{\cal{M}}} = \left[(\partial\mu_i/\partial
\mu_j);(\mu_1,\mu_2,\mu_3) = (r,t,g)\right]$ can be solved by the
expansion of the matrix elements up to order $\epsilon^{3/2}$.
When the eigenvalues $\lambda_j= A_j(b)b^{y_j}$ of
$\hat{{\cal{M}}}$ are calculated dangerous large terms of type
$b^2$ and $b^2(\mbox{ln}b)$, ($b \gg 1$)~\cite{Aharony:1974} in
the off-diagonal elements of the matrix $\hat{{\cal{M}}}$ ensure
the compensation of redundant large terms of the same type in the
diagonal elements $\hat{{\cal{M}}}_{ii}$. This compensation is
crucial for the validity of scaling for this type of critical
behavior. Such a problem does not appear in standard cases of RG
analysis~\cite{Uzunov:1993, Aharony:1974}. As in the usual
$\phi^4$--theory~\cite{Aharony:1974} the amplitudes $A_j$ depend
on the scaling factor $b$: $A_1=A_2=1+(27/13)b^2\epsilon$,
$A_3=1-(81/52)\epsilon(\mbox{ln}b)^2$. The critical exponents
$y_t=y_1$, $y_r=y_2$ and $y_g$ = $y_3$ are $b$--invariant:
\begin{equation}
\label{eq9} y_r = 2 + 10\sqrt{\frac{2\epsilon}{13}} +
\frac{197}{39}\epsilon,
\end{equation}
$y_t = y_r-18(2\epsilon/13)^{1/2}$, and $y_g = -\epsilon > 0$ for
$d < 6$. The correlation length critical exponents $\nu_{\psi} =
1/y_r$ and $\nu_M = 1/y_t$ corresponding to the fields $\psi$ and
${\mbox{\boldmath$M$}}$ are
\begin{equation}
\label{eq10} \nu_{\psi}= \frac{1}{2} -
\frac{5}{2}\sqrt{\frac{2\epsilon}{13}} + \frac{103}{156}\epsilon,
\end{equation}
\begin{equation}
\label{eq11} \nu_M = \frac{1}{2} + 2\sqrt{\frac{2\epsilon}{13}} -
\frac{5\epsilon}{156}.
\end{equation}
These exponents describe a quite particular multi-critical
behavior which differs from the numerous examples known so far.
For $d = 3$, $\nu_{\psi}$ = 0.78 which is somewhat above the usual
value $\nu \sim 0.6 \div 0.7$ near a standard phase transition of
second order~\cite{Uzunov:1993},
 but $\nu_M = 1.76$ at the same dimension ($d=3$) is unusually large.
 The fact that the Fisher's
exponent~\cite{Uzunov:1993} $\eta$ is negative for $d=3$ does not
create troubles because such cases are known in complex systems,
for example, in conventional superconductors~\cite{Halperin:1974}.
Perhaps, a direct extrapolation of the results from the present
$\epsilon$-series is not completely reliable because of the fact
that the series has been derived under the assumptions of
$\epsilon \ll 1$ and under the conditions $\epsilon^{1/2}b < 1$,
$\epsilon^{1/2}(\mbox{ln}b) \ll 1$ provided $b > 1$. These
conditions are stronger than those in the usual
$\phi^4$-theory~\cite{Uzunov:1993,Aharony:1974}. Using the known
relation~\cite{Uzunov:1993} $\gamma = (2-\eta)\nu$, the
susceptibility exponents for $d=3$ take the values $\gamma_{\psi}
= 2.06$ and $\gamma_M = 4.65$. These values exceed even those
corresponding to the Hartree approximation~\cite{Uzunov:1993}
($\gamma = 2\nu = 2$ for $d=3$) and can be easily distinguished in
experiments. Note, that here we follow the interpretation of the
asymptotic $\epsilon$-series in the way given by Lawrie et
al.~\cite{Lawrie:1987}. This point of view is quite comprehensive,
in particular, for an avoiding artificial conclusions from the RG
analysis of complex systems with competing effects, such as the
systems described by the Eq.~(1).
\begin{figure}
\begin{center}
\epsfig{file=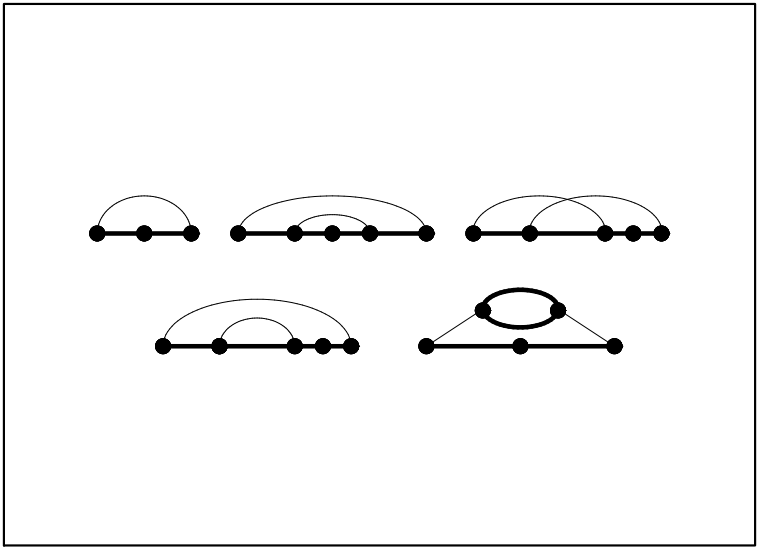,angle=-90, width=10cm}\\
\end{center}
\caption{\footnotesize Diagrams for $g^{\prime}$ of third and
fifth order in $g$. The arrows of the thick lines have been
omitted.} \label{NEW1.fig}
\end{figure}

{\em Notes about the quantum effects on the phase transitions.}
The critical behavior discussed so far may occur in a close
vicinity of finite temperature multi-critical points ($T_c=T_f>0$)
in systems possessing the symmetry of the model (1). In certain
systems, as shown in Fig.~1, this multi-critical points may occur
at $T=0$. In the quantum limit ($T\rightarrow 0$), or, more
generally, in the low-temperature limit [$T \ll \mu; \mu\equiv
(t,r);k_B=1$] the thermal wavelengths of the fields
$\mbox{\boldmath$M$}$ and $\psi$ exceed the inter-particle
interaction radius and the quantum correlations fluctuations
become essential for the critical
behavior~\cite{ShopovaPR:2003,Hertz:1976}. The quantum effects can
be considered by RG analysis of a comprehensively generalized
version of the model~(1), namely, the action ${\cal{S}}$ of the
referent quantum system. The generalized action is constructed
with the help of the substitution $(-{\cal{H}}/T) \rightarrow
S[{\mbox{\boldmath$M$}}(q),\psi(q)]$. Now the description is given
in terms of the (Bose) quantum fields $\mbox{\boldmath$M$}(q)$ and
$\psi(q)$ which depend on the $(d+1)$-dimensional vector $q =
(\omega_l, {\mbox{\boldmath$k$}})$; $\omega_l = 2\pi lT$ is the
Matsubara frequency ($\hbar=1;l = 0, \pm1,\dots$). The
${\mbox{\boldmath$k$}}$-sums in Eq.~(1) should be substituted by
respective $q$-sums and the inverse bare correlation functions ($r
+ k^2$) and ($t + k^2$) in Eq.~(1) contain additional
$\omega_l-$dependent terms, for example\cite{ShopovaPR:2003,
Hertz:1976}
\begin{equation} \label{eq12}
\langle|\psi_{\alpha}(q)|^2\rangle^{-1} = |\omega_l|+ k^2 + r.
\end{equation}
The bare correlation function $\langle|M_j(q)|\rangle^2$ contains
a term of type $|\omega_l|/k^{\theta}$, where $\theta = 1$ and
$\theta =2$ for clean and dirty itinerant ferromagnets,
respectively~\cite{Hertz:1976}. The quantum dynamics of the field
$\psi$ is described by the bare value $z=2$ of the dynamical
critical exponent $z=z_{\psi}$ whereas the quantum dynamics of the
magnetization corresponds to $z_M = 3$ (for $\theta =1$), or, to
$z_M = 4$ (for $\theta = 2$). This means that the
classical-to-quantum dimensional crossover at $T\rightarrow 0$ is
given by $ d \rightarrow (d + 2)$ and, hence, the system exhibits
a simple mean field behavior for $d \geq 4$. Just below the upper
quantum critical dimension $d_U^{(0)} =4$ the relevant quantum
effects at $T=0$ are represented by the field $\psi$ whereas the
quantum $(\omega_l$--) fluctuations of the magnetization are
relevant for $d < 3$ (clean systems), or, for even for $d < 2$
(dirty limit)~\cite{Hertz:1976}. This picture is confirmed by the
analysis of singularities of the relevant perturbation integrals.
Therefore the quantum fluctuations of the field $\psi$ have a
dominating role below spatial dimensions $ d <4$, and for
dimensions $3 < d < 4$ (clean systems), or, for $2,d<4$ in case of
dirty limit, they are the only quantum fluctuations in these
systems.

Taking into account the quantum fluctuations of the field $\psi$
and completely neglecting the $\omega_l$--dependence of the
magnetization ${\mbox{\boldmath$M$}}$, $\epsilon_0 =
(4-d)$--analysis of the generalized action ${\cal{S}}$ has been
performed within the one-loop approximation (order
$\epsilon_0^1$). In the classical limit ($r/T \ll 1$) one
re-derives the results already reported above together with an
essentially new result, namely, the value of the dynamical
exponent $z_{\psi}= 2 - (2\epsilon/13)^{1/2}$ which describes the
quantum dynamics of the field $\psi$. In the quantum limit ($r/T
\gg 1$, $T\rightarrow 0$) the static phase transition properties
are affected by the quantum fluctuations, in particular, in
isotropic systems ($n/2=m=3$). For this case, the one-loop RG
equations corresponding to $T = 0$ are not degenerate and give
definite results. The RG equation for $g$,
\begin{equation} \label{eq13} g^{\prime}=
b^{\epsilon_0/2}g\left(1 + \frac{g^2}{24\pi^3}\mbox{ln}b\right),
\end{equation}
yields two FPs: ({\em a}) a Gaussian FP ($g^{\ast} =0$), which is
unstable for $d<4$, and ({\em b}) a FP $(g^2)^{\ast} =
-12\pi^3\epsilon_0$ which is unphysical [$(g^2)^{\ast}<0$]  for
$d<4$ and unstable for $d \geq 4$. Thus the new stable critical
behavior corresponding to $T>0$ and $d<6$ disappears in the
quantum limit $T\rightarrow 0$. At the absolute zero and any
dimension $d > 0$ the $P-$driven phase transition (Fig.~1) is of
first order. This can be explained as a mere result of the limit
$T \rightarrow 0$. The only role of the quantum effects is the
creation of the new unphysical FP ({\em b}). In fact, the referent
classical system described by ${\cal{H}}$ from Eq.~(1) also looses
its stable FP (8) in the zero-temperature ({\em classical}) limit
$T\rightarrow 0$ but does not generate any new FP because of the
lack of $g^3$-term in the equation for $g^{\prime}$; see Eq.~(13).
At $T=0$ the classical system has a purely mean field
behavior~\cite{ShopovaPR:2003} which is characterized by a
Gaussian FP ($g^{\ast} = 0$) and is unstable towards
$T$--perturbations for $0<d<6$. This is a usual classical zero
temperature behavior where the quantum correlations are ignored.
For the standard $\phi^4-$ theory this picture holds for $d<4$.
One may suppose that the quantum fluctuations of the field $\psi$
are not enough to ensure a stable quantum multi-critical behavior
at $T_c=T_F=0$ and that the lack of such behavior in result of
neglecting the quantum fluctuations of $\mbox{\boldmath$M$}$. One
may try to take into account these quantum fluctuations by the
special approaches from the theory of disordered systems, where
additional expansion parameters are used to ensure the marginality
of the fluctuating modes at the same borderline dimension $d_U$
(see, e.g., Ref.~\cite{Shopova:2003}). It may be conjectured that
the techniques known from the theory of disordered systems with
extended impurities cannot be straightforwardly applied to the
present problem and, perhaps, a completely new supposition should
be introduced.

{\bf 3. Final remarks}

The present results may be of use in interpretations of recent
experiments~\cite{Pfleiderer:2002} in UGe$_2$, where the magnetic
order is uniaxial (Ising symmetry) and the experimental data, in
accord with the present consideration, indicate that the C-P phase
transition is of first order. Systems with isotropic magnetic
order are needed for an experimental test of the new
multi-critical behavior. Besides, the present investigation
exhibits several new essential problems which are a challenge to
the theory of quantum phase transitions.

{\bf Acknowledgements.} The author thanks the hospitality of
JINR-Dubna where a part of this work has been written. Financial
support by grants No.P1507 (NFSR, Sofia) and No.G5RT-CT-2002-05077
(EC, SCENET-2, Parma) is acknowledged.

\end{document}